\documentclass[12pt,a4paper]{article} 

\usepackage{amsthm}
\usepackage{cancel}
\usepackage{empheq}
\usepackage{float}
\usepackage{blindtext}
\usepackage{enumitem}
\usepackage{enumitem}
\usepackage{amsmath,latexsym,amssymb}
\usepackage{bm}
\usepackage{bbm}
\usepackage{mathtools,slashed}
\usepackage[hidelinks]{hyperref}
\usepackage{graphicx}
\usepackage[labelfont=bf]{caption}
\usepackage{xcolor}
\usepackage{appendix}
\numberwithin{equation}{section}
\definecolor{magenta}{RGB}{239, 16, 149}
\hypersetup{
 colorlinks,
 linkcolor={magenta},
 citecolor={magenta},
 urlcolor={magenta}
}
\def \N{{\mathcal N}}
\def \A{{\mathcal A}}
\def \R{{\mathcal{R}}}
\usepackage{bm}
\usepackage{bbold}
\usepackage{mathtools}
\usepackage{braket}

\usepackage[bibstyle=ieee,backend=biber]{biblatex}
\newcommand{\Tr}[1]{\mbox{Tr} #1}

\begin{document}

%###############################################################################################
%###############################################################################################
\begin{titlepage} 

\begin{center}
{\LARGE \bf Worldline path integral for the massive graviton} 
\vskip 1.2cm

Filippo Fecit$^{\,a,b}$\footnote{E-mail: filippo.fecit2@unibo.it}
\vskip 1cm

$^a${\em Dipartimento di Fisica e Astronomia ``Augusto Righi", Universit{\`a} di Bologna,\\
via Irnerio 46, I-40126 Bologna, Italy}\\[2mm]

$^b${\em INFN, Sezione di Bologna, via Irnerio 46, I-40126 Bologna, Italy}\\[2mm]
 
\end{center}
\vskip .8cm

\abstract{We compute the counterterms necessary for the renormalization of the one-loop effective action of massive gravity from a worldline perspective. This is achieved by employing the recently proposed massive $\mathcal{N}=4$ spinning particle model to describe the propagation of the massive graviton on those backgrounds that solve the Einstein equations without cosmological constant, namely on Ricci-flat manifolds, in four dimensions. The model is extended to be consistent in $D$ spacetime dimensions by relaxing the gauging of the full SO(4) R-symmetry group to a parabolic subgroup, together with the inclusion of a suitable Chern-Simons term. Then, constructing the worldline path integral on the one-dimensional torus allows for the correct calculation of the one-loop divergencies in arbitrary $D$ dimensions. Our primary contribution is the determination of the Seleey-DeWitt coefficients up to the fourth coefficient $a_3(D)$, which to our knowledge has never been reported in the literature. Its calculation is generally laborious on the quantum field theory side, as a general formula for these coefficients is not available for operators that are non-minimal in the heat kernel sense. This work illustrates the computational efficiency of worldline methods in this regard. Heat kernel coefficients characterize linearized massive gravity in a gauge-independent manner due to the on-shell condition of the background on which the graviton propagates. They could serve as a benchmark for verifying alternative approaches to massive gravity, and, for this reason, their precise expression should be known explicitly.}

\end{titlepage}
%################################################################################################################
%################################################################################################################

%################################################################################################################
\tableofcontents
%################################################################################################################

%################################################################################################################
%################################################################################################################
\section{Introduction}
Massive gravity has garnered considerable interest in theoretical physics as a compelling modification of the gravitational theory since its first formulation by Fierz and Pauli in 1939 \cite{Fierz:1939zz, Fierz:1939ix}. The investigation of this subject reached a pinnacle in recent years with the establishment of a well-defined non-linear theory known as dRGT theory \cite{deRham:2010ik, deRham:2010kj, deRham:2011rn}. Alongside these advancements, significant effort has also been devoted to investigating the quantum aspects of massive gravity, particularly in computing the one-loop divergences \cite{Buchbinder:2012wb, deRham:2013qqa}. It is not surprising that massive gravity, like general relativity, is a non-renormalizable theory, given that its Lagrangian is constructed based on the Einstein-Hilbert one, which is well known to produce diverging terms at one-loop that cannot be absorbed into the parameters of the action \cite{tHooft:1974toh, VanNieuwenhuizen:1977ca, Critchley:1978kb, Christensen:1979iy}. A viable approach to address the problem involves focusing on the theory at the linear level, namely on the Fierz-Pauli (FP) theory. In its simplest form, the FP theory is characterized by a relativistic action for a massive spin 2 particle on a flat spacetime and can be consistently extended to describe a massive graviton propagating on a curved Einstein background \cite{Buchbinder:1999ar}. This formulation of linearized massive gravity may serve as the starting point to investigate several aspects of the quantum behavior of the theory. A prominent example of such a possibility is represented by the study of the vDVZ discontinuity, which arises when taking the massless limit already at the classical level \cite{vanDam:1970vg, Zakharov:1970cc}. Interestingly, even in cases where the vDVZ discontinuity seems absent at the classical level, it has been shown to reappear at the quantum level. For instance, this has been established by calculating the one-loop graviton vacuum amplitude for a massive graviton and showing that it does not reproduce the result for the massless case in the limit $m \to 0$ \cite{Dilkes:2001av}. More recently, and most relevant to the subject of this work, the heat kernel coefficients up to the third coefficient $a_2(D)$ of linearized massive gravity have been computed in \cite{Ferrero:2023xsf}, providing the tools for an in-depth exploration of the quantum theory at the one-loop level. \\
In this work, our goal is to reproduce and extend these results by evaluating the one-loop divergences of massive gravity through a worldline approach. Recently, a worldline representation of the massive graviton has been realized relying on the so-called $O(\N)$ spinning particle models. These mechanical models have proved to be successful in achieving a first-quantized formulation of quantum field theories (QFT), describing the propagation of particles with spin $s=\frac{\N}{2}$ in four spacetime dimensions, offering an alternative perspective to conventional second-quantized field theories \cite{Berezin:1976eg, Gershun:1979fb, Howe:1988ft}. The coupling of such models to curved backgrounds faced various challenges until an approach based on BRST quantization was successfully applied for the description of Yang-Mills \cite{Dai:2008bh} and Einstein gravity \cite{Bonezzi:2018box} by means of the $\N=2$ and $\N=4$ supersymmetric spinning models, respectively. Subsequently, this approach has been applied to the case of massive gravity using the massive $\N=4$ spinning particle \cite{Fecit:2023kah}. The mass was introduced through a dimensional reduction {\it à la} Kaluza-Klein of the higher-dimensional massless model -- see \cite{Bastianelli:2005uy, Bastianelli:2014lia} for similar applications -- providing a first-quantized description for a massive graviton propagating on a curved background. To be more precise, the worldline model furnishes a first-quantized formulation of the \emph{linear} theory of massive gravity from the QFT side, which describes the propagation of massive spin $2$ particle on a non-flat background. Along the way, one finds that quantum consistency of the model requires the background metric to satisfy Einstein’s equations of motion with cosmological constant set to zero. A crucial aspect of the derivation hinges on recognizing that the associated BRST system is consistent only upon a suitable truncation of the BRST extended Hilbert space. This raises serious challenges when attempting to construct the worldline path integral on the circle. In particular, it turns out that all the unwanted states are in principle allowed to propagate in the loop. Therefore, finding the correct method to implement the projection on the massive gravity contribution becomes essential. \\
This is the issue we propose to solve in this note by adopting a similar approach to that employed for the pure gravity case in \cite{Bastianelli:2019xhi}, where the problem was addressed by modifying the measure on the moduli space, left over by the gauge-fixing, in such a way to project the full Hilbert space to the one of the spin 2 particle. This modification ensures that the graviton remains as the sole propagating degree of freedom. To test the model, we couple it to a curved Ricci-flat background and construct the path integral on the one-dimensional torus, providing a worldline representation of the one-loop effective action of massive gravity. In this way, it is possible to compute the diverging part of the effective action through the determination of the Seleey-DeWitt (SdW) coefficients. We extend previous calculations up to the fourth heat kernel coefficient, commonly referred to as $a_3(D)$, which parametrizes a class of divergences that start to appear in $D\geq 6$ dimensions and which was previously not known in the literature. Let us emphasize that the diverging terms in the one-loop effective action, evaluated on-shell, are gauge-independent and characterize unambiguously the linearized theory. Therefore, they could serve as a benchmark for verifying alternative approaches to massive gravity, and their precise expression should be known explicitly. \\

The paper is organized as follows. In section \ref{sec2} we review the free massive $\N=4$ spinning particle model and its quantization on the circle. We specifically focus on how different gaugings of the internal R-symmetry group allow for extracting different degrees of freedom. Eventually, we are able to modify the measure on the moduli space to project into the massive graviton state and extend the analysis to $D$ spacetime dimensions. In section \ref{sec3} we provide a representation of the one-loop effective action of massive gravity in the worldline formalism and compute the on-shell counterterms related to the coefficients $a_n(D)$ for $n=0,1,2,3$, providing a comparison with the existing literature. Finally, our conclusions are presented in section \ref{sec4}. 
%################################################################################################################

%################################################################################################################ 
%################################################################################################################
\section{Extracting degrees of freedom} \label{sec2}
In this section, the quantization on the circle of the free massive $\mathcal{N}=4$ spinning particle with various gaugings of the SO(4) R-symmetry is analyzed. The aim is to determine how the path integral extracts physical degrees of freedom from the Hilbert space. Once this is established, it will be possible to isolate specifically the degrees of freedom associated with the massive graviton. \\

The graded phase space of the \emph{massless} $\mathcal{N}=4$ supersymmetric worldline model consists of bosonic $(x^M,p_M)$ and fermionic $(\Xi^M_I)$ coordinates, where $M=0,\dots, D$ is a spacetime vector index and $I=1,2,3,4$ is a SO(4) internal index. The target space is a $(D+1)-$dimensional Minkowski space $\mathcal{M}_{D+1}$ and $t$ denotes a parameter that labels positions along the worldline, which is embedded in spacetime by the functions $x^M(t)$. The phase space action is given by
\begin{equation} \label{act}
S=\int d t \left[p_M\dot x^{M}+\frac{i}{2}\,\Xi_{M}^I \dot\Xi^{M}_I-\frac{e}{2}\,\mathcal{H}-i \, \mathcal{X}^I Q_I\right]\ ,
\end{equation}
with $(e, \mathcal{X})$ being a one-dimensional supergravity multiplet enforcing the first-class constraints $(\mathcal{H}, Q)$ that generate through Poisson brackets the $\N=4$ superalgebra on the worldline
\begin{equation}
\{Q_I, Q_J\}_{\mathrm{PB}}=-2i\,\delta_{IJ}\,\mathcal{H}\ ,\quad \{Q_I, \mathcal{H}\}_{\mathrm{PB}}=0\ .
\end{equation}
The latter algebra is computed by using the graded Poisson brackets of the phase space coordinates $\{x^M, p_N\}_{\mathrm{PB}}=\delta^M_N$ and $\{\Xi_{M}^I, \Xi^{N}_J\}_{\mathrm{PB}}=-i\delta_M^N \delta^I_J$, fixed by the symplectic term of the action \cite{new-book}. The constraints $\mathcal{H}:=p^2$ and $Q_I:=\Xi_I^M p_M$ must be introduced to ensure the mass-shell condition and to remove unphysical degrees of freedom, eliminating negative norm states. From the latter higher-dimensional theory, the lower-dimensional \emph{massive} model is derived as follows. First, it is convenient to take complex combinations of the original four real Grassman variables $\Xi^M_I(t)$ and the four gravitinos $\mathcal{X}^I(t)$ as follows ($i=1,2$)
\begin{align}
\begin{split} \label{complex}
\xi^M_i&:=\tfrac{1}{\sqrt2}(\Xi^M_i+i\,\Xi^M_{i+2})\ , \quad \bar\xi^{M i}:=\tfrac{1}{\sqrt2}(\Xi^M_i-i\,\Xi^M_{i+2})\ , \\
\chi_i&:=\tfrac{1}{\sqrt2}(\mathcal{X}_i+i\,\mathcal{X}_{i+2})\ , \quad \bar\chi^{i}:=\tfrac{1}{\sqrt2}(\mathcal{X}_i-i\,\mathcal{X}_{i+2})\ .
\end{split}
\end{align}
Then, employing the so-called Scherk-Schwarz mechanism \cite{Scherk:1979zr}, the model is dimensionally reduced on a flat spacetime of the form $\mathcal{M}_D \times S^1$. In practice, one gauges the compact direction $x^D$, corresponding to $S^1$, by imposing the first-class constraint $p_D=m$ while setting
\begin{align} \label{DEC}
x^M = (x^\mu , x^D)\ , \quad p_M = (p_\mu , p_D)\ , \quad \xi^M_i=(\psi^\mu_i,\theta_i)\ , \quad \bar \xi^{M i}=(\bar \psi^{\mu i},\bar \theta^i)\ .
\end{align}
We refer the reader to \cite{Bastianelli:2014lia, Fecit:2023kah} for further details on the derivation. The worldline phase space action of the massive $\mathcal{N}=4$ spinning particle is given by
\begin{equation} \label{action}
S=\int d t \left[p_\mu\dot x^{\mu}+i\bar\psi_{\mu}\cdot\dot\psi^{\mu}+i\bar \theta \cdot \dot \theta-\frac{e}{2}\,H-i\chi \cdot \bar q-i\bar \chi \cdot q\right]\ , 
\end{equation}
where a dot indicates a contraction of the internal indices. The worldline action \eqref{action} displays many symmetries. Specifically, the local symmetries are worldline reparametrizations generated by the Hamiltonian $(H)$ and four worldline supersymmetries generated by the supercharges $(q, \bar q)$, where
\begin{equation} \label{const}
H:=p^\mu p_\mu+m^2\ , \quad q_i:=\psi_i^\mu \, p_\mu+m\theta_i\ , \quad \bar q^i:=\bar \psi^{i\mu} \, p_\mu+m\bar \theta^i\ .
\end{equation}
Let us stress that their presence is \emph{essential} to describe relativistic massive particles in target space as they ensure unitarity: for this very reason said symmetries have been made local. The aforementioned worldline supergravity multiplet $(e,\chi,\bar \chi)$ acts as a set of Lagrange multipliers for the first-class constraints \eqref{const}. \\
Upon quantization, the worldline coordinates obey the following (anti)commutation relations fixed by their classical Poisson brackets
\begin{equation}
[x^\mu, p_\nu]=i\,\delta^\mu_\nu\ ,\quad \{\bar \psi^{\mu i}, \psi^{\nu}_j\}=\delta^i_j\,\eta^{\mu\nu}\ , \quad \{\bar \theta^i, \theta_j\}=\delta^i_j\ .
\end{equation}
By choosing a Fock vacuum annihilated by $(\bar \psi^i_\mu, \bar \theta^i)$, a generic state $\ket{\Omega}$ in the Hilbert space can be identified with the wavefunctions
\begin{equation}\label{wavefunction}
\Omega(x,\psi_i,\theta_i)=\sum_{n_1,n_2=0}^D \; \sum_{m_1,m_2=0}^1\Omega_{\mu[n_1]\vert \nu[n_2]}(x)\,\psi_1^{\mu_1} \dots \psi_1^{\mu_{n_1}}\theta_1^{m_1}\,\psi_2^{\nu_1} \dots \psi_2^{\nu_{n_2}}\theta_2^{m_2}\ ,
\end{equation}
namely a collection of tensor fields with the symmetries of $(n_1,n_2)$ bi-forms. We used the condensed notation for antisymmetrized indices $\mu[n]:=[\mu_1 \dots \mu_n]$ and a vertical bar to separate indices with no symmetry relations. \\

The spectrum \eqref{wavefunction} contains way too many states, which is reflected on the corresponding BRST system, found to be consistent only upon a suitable truncation of the BRST Hilbert space. While this works for BRST cohomology, the problem reappears at the one-loop level, since in principle all the unwanted states may propagate in the loop. Our task is thus to find a way to implement the projection on the massive gravity sector. The first step in this direction consists of restricting the spectrum to the $n_1+m_1=1$, $n_2+m_2=1$ subspace. Indeed, it has been shown \cite{Bastianelli:2019xhi} that the massless sector contains the massless NS-NS spectrum of closed string theory, and corresponds to the level $(n_1,n_2\,\vert\,m_1,m_2)=(1,1\,\vert\,0,0)$ fields $\Omega_{\mu\vert\nu}$. The latter decomposes into a graviton, an antisymmetric Kalb-Ramond two-form, and a dilaton 
\begin{equation}
\Omega_{\mu\vert\nu}(x)=h_{\mu\nu}(x)+B_{\mu\nu}(x)+\delta_{\mu\nu} \, \phi(x)\ ,
\end{equation}
with the graviton identified with the symmetric and traceless component. Due to the mass improvement, we anticipate the spectrum in the level $(m_1,m_2)\neq(0,0)$ to include also the associated St\"uckelberg fields $(\Omega_\mu, \Omega_\nu, \Omega)$, specifically two massless vector fields and a scalar \cite{Stueckelberg:1957zz, Boulanger:2018dau}, allowing for the propagation of the massive graviton and the massive Kalb-Ramond degrees of freedom alongside the dilaton. This will be confirmed through the analysis of the worldline path integral. \\

In order to implement the aforementioned projection, the R-symmetry of the model has to be appropriately exploited. Notice that the action \eqref{act} has a manifest global SO(4) symmetry that rotates the fermions, generated by the fermion bilinears
\begin{equation}
\mathrm{J}_{IJ}:=i\,\Xi_{I}^M \, \Xi_{J M}\ ,
\end{equation}
while in the complex basis \eqref{complex} and with the decomposition \eqref{DEC} only the subgroup U(2) $\subset$ SO(4) is kept manifest as rigid symmetry of the action \eqref{action}. As a result, the SO(4) generators split as $\mathrm{J}_{IJ} \sim (J_i^j,K^{ij},G_{ij})$ and their explicit realization is\footnote{See appendix B of \cite{Bonezzi:2018box} for a detailed derivation.}
\begin{align} 
\begin{split} \label{gen}
J_i^j &:=\psi_i \cdot {\bar{\psi}}^j+\theta_i \, \bar{\theta}^j\ , \\
K^{ij} &:= \bar{\psi}^i \cdot \bar{\psi}^j+\bar{\theta}^i \, \bar{\theta}^j\ , \\
G_{ij} &:= \psi_i\cdot \psi_j+\theta_i \, \theta_j\ ,
\end{split}
\end{align}
where we used a dot $\cdot$ to indicate contraction on spacetime indices. $K^{ij}$ is the so-called trace operator while $G_{ij}$ implements the insertion of the metric; they both vanish for $i=j$ \cite{Bastianelli:2014lia, Bonezzi:2018box}. On the other hand, $J_i^j$ generate the U(2) subgroup, and upon quantization become quantum operators\footnote{We used hats to stress that the expression refers to operators in that given order; however, throughout the text, we will often avoid the use of hats for quantum operators, so as not to burden the notation further.}
\begin{equation}
\hat J_i^j = \hat \psi_i \cdot \hat {\bar{\psi}}^j+\hat \theta_i \, \hat{\bar{\theta}}^j-\tfrac{D+1}{2}\,\delta_i^j\ ,
\end{equation}
where we employed the Weyl ordering to resolve ambiguities, matching the path integral regularization used. It is worth noticing that the shift $\tfrac{D+1}{2}$ is a quantum ordering effect. \\
The crucial point is that it is possible, although not strictly necessary, to make the R-symmetry local. Differently from worldline supersymmetries, the R-symmetry has not to be gauged for unitarity; however, it can be used to perform algebraic projections on the desired spectrum by appropriately gauging only a specific subgroup $\R$ $\subset$ SO(4). At the level of the worldline action, this corresponds to introducing the appropriate worldline gauge fields\footnote{Here we use $\R$ as a mere label.} $a^{\mathcal{R}}(t)$, by the insertion of a term
\begin{equation}
S_{\R}=-\int d t \; a^{\mathcal{R}}\,\mathrm{J}_{\mathcal{R}}\ ,
\end{equation}
acting as Lagrange multipliers for the classical constraints $\mathrm{J}_{\mathcal{R}}$. In the following, we seek to unveil the massive gravity content with different choices of the $\R$ subgroup, which extracts degrees of freedom from the worldline path integral. 

%----------------------------------------------------------
\subsubsection*{Setting up the worldloop path integral}
%----------------------------------------------------------
The path integral on worldlines with the topology of a circle -- dubbed ``worldloop'' -- is defined as
\begin{equation} \label{1}
\Gamma = \int_{S^1}
\frac{DG\,DX}{\mathrm{Vol(Gauge)}}\, \mathrm{e}^{-S_{\mathrm{E}}[X,G]}\ ,
\end{equation}
where we denoted the worldline gauge fields $G=\left( e, \chi, \bar{\chi},a\right)$ and the coordinates with supersymmetric partners $X=\left(x,\psi,\bar{\psi},\theta,\bar \theta\right)$. The action appearing in \eqref{1} is the action in Euclidean configuration space ($S_{\mathrm{E}}=-iS$), obtained by a Wick-rotation to Euclidean time $t \to -i\tau$ accompanied by the Wick rotations of the gauge fields $a_{\R}\to -ia_{\R}$, just as done in \cite{Bastianelli:2007pv} for general $\N$. From now on we will drop the subscript on $S_{\mathrm{E}}$ as no confusion should arise. The overcounting from summing over gauge equivalent configurations, which causes the path integral to diverge, is formally taken into account by dividing by the volume of the gauge group. To regularize the path integral one has to follow a gauge-fixing procedure. We use the Faddeev-Popov ($\Phi\Pi$) method to extract the volume of the gauge group and to gauge-fix completely the supergravity multiplet up to some moduli while evaluating the determinants stemming from the associated $\Phi\Pi$ ghosts \cite{Faddeev:1967fc}, as we will outline in the following. \\

The einbein is gauge-fixed to a constant, namely $e(T)=2T$, where $T$ is often called ``Schwinger proper time", while the gravitinos $(\chi, \bar{\chi})$ are antiperiodic and gauge-fixed to zero, leaving no additional moduli. The path integral \eqref{1} multiplied by $-\tfrac{1}{2}$ corresponds to the QFT effective action and can be recast in the following form
\begin{equation} \label{euc}
\Gamma = -\frac{1}{2}\int_{0}^{\infty}\frac{dT}{T}\,\mathrm{e}^{-m^2T} \,Z_{\R}(T)\ ,
\end{equation} 
%In the next section, we will see that \eqref{euc} corresponds to the one-loop effective action $\Gamma[g_{\mu\nu}]$ of massive gravity coupling the worldline model to a curved background metric $g_{\mu\nu}(x)$. In that regard, the overall normalization $-\tfrac{1}{2}$ is inserted to match QFT results, whereas
where the integration over the Schwinger proper time arises from the gauge-fixing of the einbein $e$. The explicit expression of the density $Z_{\R}(T)$ and of the gauge-fixing conditions for the worldline gauge fields $a_{\R}$ depends on the subset $\R$ being gauged. We start our analysis with the simplest case of gauging the $\R=$ U(1) $\times$ U(1) subgroup to set the grounds for the general case. 

%----------------------------------------------------------
\subsection{Gauging of the U(1) $\times$ U(1) subgroup}
%----------------------------------------------------------
The U(1) $\times$ U(1) subgroup corresponds to the generators
\begin{equation} \label{U(1)}
J_i:=\psi_i \cdot \bar{\psi}^i + \theta_i \, \bar{\theta}^i \quad \text{($i$ not summed)}\ .
\end{equation}
The gauging of the U(1) $\times$ U(1) subgroup is realized by means of two abelian worldline gauge fields $a_i(t)$. At the level of the worldline action, this corresponds to
\begin{equation}
S_{\mathrm{U(1)}\times \mathrm{U(1)}}=-\int d t \; a^i (J_i -q_i )\ ,
\end{equation}
where two independent Chern-Simons (CS) couplings $q_i=\tfrac{3-D}{2}$, that convert the classical constraints $(J_i -q_i )$ into the operatorial constraints $(\hat N_i -2)$,\footnote{With $\hat N_i:=\hat N_{\psi_i} + \hat N_{\theta_i}=\hat \psi_i \cdot \hat{\bar{\psi}}^i+\hat \theta_i \, \hat{\bar{\theta}}^i$ being the number operators counting the number of fermionic oscillators with a fixed flavor index. The U(1) $\times$ U(1) generators explicitly read $\hat J_i:=\hat N_i-\tfrac{D+1}{2}$.} have been included to project on the gravity sector. The whole supergravity multiplet is gauge-fixed as 
\begin{equation}
\tilde G=\left(T, 0, 0,\tilde a_{i}\right) \quad \text{with} \quad \tilde a_{i}:= \left(\begin{array}{c} \alpha \\ \beta \\
\end{array}\right)\ ,
\end{equation}
where $\alpha, \beta \in [0,2\pi]$ are two angles representing additional moduli related to the gauge fields. Inserting the $\Phi\Pi$ determinants to eliminate the volume of the gauge group, and setting appropriately the overall normalization, the partition function in \eqref{euc} explicitly reads
\begin{equation} \label{path}
Z_{\mathrm{U(1)}\times \mathrm{U(1)}}(T)=\int_0^{2\pi}\frac{d\alpha}{2\pi}\int_0^{2\pi}\frac{d\beta}{2\pi}\,\left(2\cos\tfrac{\alpha}{2}\right)^{-2}\left(2\cos\tfrac{\beta}{2}\right)^{-2} \Tr\left[\mathrm{e}^{-T\hat H}\mathrm{e}^{i\alpha(\hat N_{\psi_1}+\hat N_{\theta_1}-2)+i\beta(\hat N_{\psi_2}+\hat N_{\theta_2}-2)}\right]\ .
\end{equation}
A few comments are in order. The two gauge fields $a_i$ produce the integration over the angular moduli $(\alpha,\beta)$, while the SUSY ghosts account for the cosine factors. The path integral over the ``matter" sector
\begin{equation}
\int_{_{\rm PBC}}\hskip-.4cm{ D}x\int_{_{\rm ABC}}\hskip-.4cm D\bar{\psi}D\psi\int_{_{\rm ABC}}\hskip-.4cm D\bar{\theta}D\theta \, \mathrm{e}^{-S_{\mathrm{gf}}[X,\tilde G]}
\end{equation}
has been put into an operatorial form as a trace -- with $\hat H = \hat p^2$ for the free theory and with $\hat N$ being the operators counting the number of oscillators with a fixed flavor index -- over the Hilbert space consisting of differential forms of arbitrary degree, contained in the Taylor coefficients of the wavefunctions $\Omega(x,\psi_i,\theta_i)$. The path integration over bosonic variables is evaluated by fixing periodic boundary conditions (PBC), while the fermionic path integral is performed by choosing antiperiodic boundary conditions (ABC) on each flavor of fermionic fields. Finally, the gauge-fixed action reads
\begin{equation} \label{actiongf}
S_{\mathrm{gf}}[X,\tilde G]=\int d\tau \left[ \frac{1}{4T}\dot{x}^{\mu}\dot{x}_{\mu}+m^2T+\sum_{i=1,2}\bar{\psi}_i^{\mu}(\partial_\tau +i \tilde a_{i}){\psi}^i_{\mu}+\sum_{i=1,2}\bar{\theta}_{i}(\partial_\tau +i \tilde a_{i}){\theta}^i +i\sum_{i=1,2}\tilde a^i \, q_i\right]\ .
\end{equation}
Having set up the worldloop, we are in the position of analyzing the degrees of freedom of the wordline model. Firstly, note that the Hilbert space can be decomposed in terms of the eigenvalues $(n_1,n_2)$ and $(m_1,m_2)$ of the pairs of number operators $(\hat N_{\psi_1},\hat N_{\psi_2})$ and $(\hat N_{\theta_1},\hat N_{\theta_2})$ respectively. Consequently, the trace can be decomposed in terms of 
\begin{align}
\begin{split} \label{N}
N_1 &=n_1+m_1\ , \\
N_2 &=n_2+m_2\ ,
\end{split}
\end{align}
reproducing the double grading of the \emph{massless} $\mathcal{N}=4$ spinning particle \cite{Bastianelli:2019xhi}
\begin{equation}
t_{N_1,N_2}(T):=\Tr_{N_1,N_2}\left[\mathrm{e}^{-T\hat H}\right]\ .
\end{equation}
The path integral \eqref{path}, using the Wilson line variables $z:= \mathrm{e}^{i\alpha}$ and $\omega:= \mathrm{e}^{i\beta}$, becomes\footnote{The modular integration is performed over the circle $|z|=1$, with the singular point $z=-1$ pushed out of the contour, the same goes for $\omega$. Discussion on the regulated contour of integration for the modular parameters can be found in \cite{Bastianelli:2005vk}.}
\begin{equation}
Z_{\mathrm{U(1)}\times \mathrm{U(1)}}(T)=\oint \frac{dz}{2\pi i z}\oint \frac{d\omega}{2\pi i \omega}\,\frac{z}{(z+1)^2}\frac{\omega}{(\omega+1)^2} \sum_{N_1,N_2} t_{N_1,N_2}(T) \, z^{N_1-2}\omega^{N_2-2}\ .
\end{equation}
Then, we can trace back the contributions from the $\theta$s through the following identification 
\begin{equation} \label{dec}
\sum_{N_1,N_2} t_{N_1,N_2} \equiv \sum_{n_1,n_2=0}^D \; \sum_{m_1,m_2=0}^1 \mathbbmss{T}\,_{n_1,n_2}^{m_1,m_2}\ ,
\end{equation}
where $\mathbbmss{T}\,_{n_1,n_2}^{m_1,m_2}$ denotes the trace restricted to some specific eigenvalues $(n_1,n_2\,\vert\,m_1,m_2)$ of the fermionic number operators. Let us highlight how the ``massless to massive decomposition" works. Upon modular integration, the massless partition function includes only the following contributions
\begin{equation}
Z_{\mathrm{U(1)}\times \mathrm{U(1)}}=t_{1,1}-2\,t_{1,0}-2\,t_{0,1}+4\,t_{0,0}\ ,
\end{equation}
while the other possible values of $(N_1, N_2)$ yield zero. Keeping in mind the decomposition \eqref{N}--\eqref{dec}, in the following denoted with arrows, we get the massive improvements listed below
\begin{align}
\begin{split}
t_{1,1} &\xlongrightarrow{m} \quad \mathbbmss{T}\,_{1,1}^{0,0}+\mathbbmss{T}\,_{1,0}^{0,1}+\mathbbmss{T}\,_{0,1}^{1,0}+\mathbbmss{T}\,_{0,0}^{1,1}\ , \\[2mm]
t_{1,0} &\xlongrightarrow{m} \quad \mathbbmss{T}\,_{1,0}^{0,0} +\mathbbmss{T}\,_{0,0}^{1,0}\ ,\\[2mm]
t_{0,1} &\xlongrightarrow{m} \quad \mathbbmss{T}\,_{0,1}^{0,0} +\mathbbmss{T}\,_{0,0}^{0,1}\ ,\\[2mm]
t_{0,0} &\xlongrightarrow{m} \quad \mathbbmss{T}\,_{0,0}^{0,0}\ .
\end{split}
\end{align}
The above partition function can then be decomposed into its irreducible spacetime components. The degrees of freedom for the free theory are given by
\begin{equation}
\mathbbmss{T}\,_{n_1,n_2}^{m_1,m_2}=\frac{1}{(4\pi T)^{D/2}}\binom{D}{n_1}\binom{D}{n_2}\ ,
\end{equation}
where the factor $\left(4\pi T\right)^{-D/2}$ corresponds to the free particle position and will be omitted in the following. On the other hand, the binomials count the number of DOFs and correspond to the transverse polarizations of the tensor $\Omega_{\mu\vert\nu}$, yielding
\begin{equation}
Z_{\mathrm{U(1)}\times \mathrm{U(1)}}= Z_{h_{\mu\nu}}+Z_{A_{\mu}}+Z_\varphi+Z_{B_{\mu\nu}}+Z_{C_{\mu}}+Z_\phi=(D-1)^2\ , 
\end{equation}
where 
\begin{equation}
Z_{h_{\mu\nu}}+Z_{A_{\mu}}+Z_\varphi=\frac{(D+1)(D-2)}{2}
\end{equation}
corresponds to a \emph{massive} graviton, i.e. a massless graviton $Z_{h_{\mu\nu}}$ along with the two St\"uckelberg fields, a massless vector field $Z_{A_{\mu}}$ and a scalar field $Z_\varphi$\footnote{Eventually, one could identify them as a single \emph{massive} vector St\"uckelberg carrying $D-1$ degrees of freedom.}
\begin{align}
Z_{h_{\mu\nu}} &\equiv \mathbbmss{T}\,_{1,1}^{0,0}-\mathbbmss{T}\,_{2,0}^{0,0}-2 \, \mathbbmss{T}\,_{1,0}^{0,0}=\tfrac{D(D-3)}{2}\ , \\
Z_{A_{\mu}} &\equiv\mathbbmss{T}\,_{1,0}^{0,1}-2 \, \mathbbmss{T}\,_{0,0}^{1,0}=D-2\ ,\\
Z_\varphi &\equiv \mathbbmss{T}\,_{0,0}^{1,1}=1\ ,
\end{align}
while 
\begin{equation}
Z_{B_{\mu\nu}}+Z_{C_{\mu}} =\frac{(D-1)(D-2)}{2}
\end{equation}
is the massive Kalb-Ramond field, corresponding to the massless contribution plus the massless vector St\"uckelberg
\begin{align}
Z_{B_{\mu\nu}} &\equiv \mathbbmss{T}\,_{2,0}^{0,0}-2 \, \mathbbmss{T}\,_{0,1}^{0,0}+3 \, \mathbbmss{T}\,_{0,0}^{0,0}=\tfrac{(D-2)(D-3)}{2}\ ,\\
Z_{C_{\mu}} &\equiv \mathbbmss{T}\,_{0,1}^{1,0}-\mathbbmss{T}\,_{0,0}^{0,1}=D-2\ ,
\end{align}
and with
\begin{equation}
Z_\phi \equiv \mathbbmss{T}\,_{0,0}^{0,0}=1
\end{equation}
being the dilaton. One can check that the $(m_1, m_2)=(0,0)$ sector correctly reproduces the massless spectrum of the Hilbert space contained in the $\mathcal{N}=4$ spinning particle model, which coincides with the massless NS-NS sector of closed strings, while the $(m_1,m_2) \neq (0,0)$ sector corresponds to the massive improvements, namely the associated St\"uckelberg fields.

%----------------------------------------------------------
\subsection{Full SO(4) group gauging}
%----------------------------------------------------------
In the following, we gauge the entire R-symmetry group. The analysis will show that this choice produces not only the massive graviton in the spectrum but also some unwanted contributions. One has to find a way to project the latter away to finally construct a worldline path integral specifically for the massive graviton. \\

The gauging of the full set of generators is achieved through a one-dimensional Yang-Mills field $a^{IJ}(t)$ acting as a Lagrange multiplier in \eqref{action}. Explicitly
\begin{equation}
S_{\mathrm{SO(4)}}=-\frac{1}{2}\int d t \; a^{IJ} \mathrm{J}_{IJ}\ ,
\end{equation}
or, taking into account the splitting \eqref{gen},
\begin{equation}
S_{\mathrm{SO(4)}}= -\int d t \left( a_i^j J_j^i+\frac12 a_{ij} K^{ij}+\frac12 a^{ij} G_{ij}\right)\ .
\end{equation}
Note that, contrary to the previous case, there is no room for Chern-Simons couplings here. It is necessary then to restrict the analysis to $D=3$ spacetime dimensions at first, to correctly repdouce a graviton state. This is related to the BRST quantization performed in \cite{Fecit:2023kah}, where it was discussed how the physical wavefunction of the spinning particle lies in the kernel of the operator $\hat{\mathcal{J}}_i^i$ -- which is the quantum operator corresponding to \eqref{U(1)} including the contribution from the BRST bosonic superghosts -- only in three spacetime dimensions. Even if it has been shown how to reproduce a first-quantized massive gravity theory in four spacetime dimensions by demanding the physical states to have a fixed U(1) $\times$ U(1) charge of $-\tfrac{1}{2}$, in the present case the latter condition could be satisfied only upon adding a Chern-Simons term, which is not possible when the full gauging is considered. At the end of the analysis, it shall then be discussed how to overcome this obstruction. \\ 

The starting point is the path integral \eqref{euc}, with the gauge-fixing $\tilde G=\left(T, 0, 0,\tilde a_{i}^{j}\right)$ producing now the following action
\begin{equation} \label{azione}
S_{\mathrm{gf}}[X,\tilde G]=\int d\tau \left[ \frac{1}{4T}\dot{x}^{\mu}\dot{x}_{\mu}+m^2T+\bar{\psi}^{i \mu}\left(\delta_{ij} \partial_\tau +i \tilde a_{ij}\right){\psi}^j_{\mu}+\bar{\theta}^{i}\left(\delta_{ij} \partial_\tau +i \tilde a_{ij}\right){\theta}^j\right]\ , 
\end{equation}
with
\begin{equation}
\tilde a_{i}^{j}:=\left(\begin{array}{cc} \alpha & 0 \\ 0 & \beta \\
\end{array}\right)\ .
\end{equation}
In the present case, one has to introduce the non-abelian $\Phi\Pi$ ghosts associated with the whole SO(4) group, whose path integration modifies the path integral \eqref{path} with the inclusion of the sine factors previously calculated in \cite{Bastianelli:2007pv}. 
%\begin{equation} \label{path2}
%Z_{\mathrm{SO(4)}}(T)=\int_0^{2\pi}\frac{d\alpha}{2\pi}\int_0^{2\pi}\frac{d\beta}{2\pi} \, \frac{1}{4}\left(2i\,\sin\tfrac{\alpha+\beta}{2}\right)^2\left(2i\,\sin\tfrac{\alpha-\beta}{2}\right)^2 \, \left(2\cos\tfrac{\alpha}{2}\right)^{-2}\left(2\cos\tfrac{\beta}{2}\right)^{-2} \, \int_{S^1}DX\, \mathrm{e}^{-S_{\mathrm{gf}}[X,\tilde G]}\ .
%\end{equation}
In terms of Wilson variables the path integral explicitly reads\footnote{Note that we factored out the $\tfrac{1}{4}$ also stemming from the path integral over the SO(4) $\Phi\Pi$ ghosts for simplicity.}
\begin{equation}
Z_{\mathrm{SO(4)}}(T)=\frac{1}{4}\oint \frac{dz}{2\pi i z}\oint \frac{d\omega}{2\pi i \omega}\,\frac{z}{(z+1)^2}\frac{\omega}{(\omega+1)^2} \, p(z,\omega)\sum_{N_1,N_2} t_{N_1,N_2}(T) \,z^{N_1-2}\omega^{N_2-2}\ ,
\end{equation}
where $p(z,\omega)$ is the measure on the moduli space arising from the full gauging and reads
\begin{equation} \label{p}
p(z,\omega):= 4-2z\omega-2\left(\frac{z}{\omega}+\frac{\omega}{z}\right)+z^2+\omega^2-\frac{2}{z\omega}+\frac{1}{z^2}+\frac{1}{\omega^2}\ .
\end{equation}
To unveil the projection of the various components of the full measure, it is possible to follow the ``massless to massive decomposition" described in the previous subsection. Take as an illustrative example the first monomial of \eqref{p}. It yields the following contribution upon modular integration
\begin{align}
4 \; &\xlongrightarrow{\oint\oint} \quad t_{1,1}-2\,t_{1,0}-2\,t_{0,1}+4\,t_{0,0} \\[2mm]
&\xlongrightarrow{m} \quad \mathbbmss{T}\,_{1,1}^{0,0}+\mathbbmss{T}\,_{1,0}^{0,1}+\mathbbmss{T}\,_{0,1}^{1,0}+\mathbbmss{T}\,_{0,0}^{1,1}-2\mathbbmss{T}\,_{1,0}^{0,0}-2\mathbbmss{T}\,_{0,1}^{0,0} -2\mathbbmss{T}\,_{0,0}^{1,0}-2\mathbbmss{T}\,_{0,0}^{0,1}+4\mathbbmss{T}\,_{0,0}^{0,0} \nonumber \\[2mm]
&\equiv \quad \;\;\; Z_{\mathrm{U(1)}\times \mathrm{U(1)}}\ . \nonumber
\end{align}
The first arrow implies a modular integration, while the second one denotes the massless to massive decomposition as previously discussed. As for the remaining monomials, the outcome is as follows:
\begin{align}
-2z\omega \; &\xlongrightarrow{\oint\oint} \quad -\tfrac{1}{2}\,t_{0,0} \; \xlongrightarrow{m} \quad -\tfrac{1}{2}\, \mathbbmss{T}\,_{0,0}^{0,0} \; \equiv \quad -\tfrac12\, Z_\phi\ , \\[3mm]
-2\left(\frac{z}{\omega}+\frac{\omega}{z}\right) \;&\xlongrightarrow{\oint\oint} \quad -(t_{2,0}-2\,t_{1,0}+3\,t_{0,0}) \\[2mm]
&\xlongrightarrow{m} \quad -\left( \mathbbmss{T}\,_{2,0}^{0,0}+\mathbbmss{T}\,_{1,0}^{1,0}-2\,\mathbbmss{T}\,_{1,0}^{0,0}-2\,\mathbbmss{T}\,_{0,0}^{1,0}+3\mathbbmss{T}\,_{0,0}^{0,0}\right) \nonumber \\[2mm]
&\equiv \quad \;\;\; -\left( Z_{B_{\mu\nu}} + Z_{C_{\mu}} \right)\ , \nonumber \\[3mm]
z^2+\omega^2\;&\xlongrightarrow{\oint\oint} \quad 0\ ,
\end{align}
for the first half of $p(z,\omega)$. It is evident that, as for the massless case, the above monomials are enough to construct a measure capable of projecting only into the massive gravity sector. We list here the contributions stemming from the other monomials for completeness: we have
\begin{align}
-\frac{2}{z\omega}\;&\xlongrightarrow{\oint\oint} \quad -\tfrac12\, t_{2,2}-2\,t_{1,1}-\tfrac92\,t_{0,0}+2\,t_{2,1}-3\,t_{2,0}+6\,t_{1,0} \\[2mm]
&\xlongrightarrow{m} \quad -\tfrac12\,\mathbbmss{T}\,_{2,2}^{0,0}-\tfrac12\,\mathbbmss{T}\,_{2,1}^{0,1}-\tfrac12\,\mathbbmss{T}\,_{1,2}^{1,0}-\tfrac12\,\mathbbmss{T}\,_{1,1}^{1,1} 
-2\,\mathbbmss{T}\,_{1,1}^{0,0}-2\,\mathbbmss{T}\,_{1,0}^{0,1}-2\,\mathbbmss{T}\,_{0,1}^{1,0}-2\,\mathbbmss{T}\,_{0,0}^{1,1} \nonumber \\[1mm]
&\phantom{\xlongrightarrow{m}} \; \quad -\tfrac92\,\mathbbmss{T}\,_{0,0}^{0,0}
+2\,\mathbbmss{T}\,_{2,1}^{0,0}+2\,\mathbbmss{T}\,_{2,0}^{0,1}+2\,\mathbbmss{T}\,_{1,1}^{1,0}+2\,\mathbbmss{T}\,_{1,0}^{1,1}
-3\,\mathbbmss{T}\,_{2,0}^{0,0}-3\,\mathbbmss{T}\,_{1,0}^{1,0}
+6\,\mathbbmss{T}\,_{1,0}^{0,0}+6\,\mathbbmss{T}\,_{0,0}^{1,0} \nonumber \\[2mm]
&\equiv \quad \;\;\; -\tfrac12\,Z_{\A_{2,2}}-Z_{\A_{2,1}}-\tfrac12\,Z_{\A_{1,1}}\ , \nonumber \\[3mm]
\frac{1}{z^2}+\frac{1}{\omega^2}\;&\xlongrightarrow{\oint\oint} \quad \tfrac12\, t_{3,1}-\,t_{2,1}+\tfrac32\,t_{1,1}-5\,t_{1,0}-\,t_{3,0}+2\,t_{2,0}+4\,t_{0,0} \\[2mm]
&\xlongrightarrow{m} \quad +\tfrac12\,\mathbbmss{T}\,_{3,1}^{0,0}+\tfrac12\,\mathbbmss{T}\,_{3,0}^{0,1}+\tfrac12\,\mathbbmss{T}\,_{2,1}^{1,0}+\tfrac12\,\mathbbmss{T}\,_{2,0}^{1,1}
-\,\mathbbmss{T}\,_{2,1}^{0,0}-\,\mathbbmss{T}\,_{2,0}^{0,1}-\,\mathbbmss{T}\,_{1,1}^{1,0}-\,\mathbbmss{T}\,_{1,0}^{1,1} \nonumber \\[1mm]
&\phantom{\xlongrightarrow{m}} \; \quad +\tfrac32\,\mathbbmss{T}\,_{1,1}^{0,0}+\tfrac32\,\mathbbmss{T}\,_{1,0}^{0,1}+\tfrac32\,\mathbbmss{T}\,_{0,1}^{1,0}+\tfrac32\,\mathbbmss{T}\,_{0,0}^{1,1}
-5\,\mathbbmss{T}\,_{1,0}^{0,0}-5\,\mathbbmss{T}\,_{0,0}^{1,0}
-\,\mathbbmss{T}\,_{3,0}^{0,0}-\,\mathbbmss{T}\,_{2,0}^{1,0} \nonumber \\[1mm]
&\phantom{\xlongrightarrow{m}} \; \quad +2\,\mathbbmss{T}\,_{2,0}^{0,0}+2\,\mathbbmss{T}\,_{1,0}^{1,0}
+4\,\mathbbmss{T}\,_{0,0}^{0,0} \nonumber \\[2mm]
&\equiv \quad \;\;\; \tfrac12\,Z_{\A_{3,1}}+\tfrac12\,Z_{\A_{3,0}}+\tfrac12\,Z_{\A_{2,1}}+\tfrac12\,Z_{\A_{2,0}}\ . \nonumber
\end{align}
The above contributions can be interpreted in terms of bi-forms $\A_{p,q}$ with corresponding partition functions 
\begin{equation}
Z_{\A_{p,q}}=\sum_{k,l=0}^{p,q}(-)^{k+l}(k+1)(l+1)\,t_{p-k,q-l} 
\end{equation}
previously analyzed in \cite{Bastianelli:2019xhi}. Indeed, they coincide with the same topological contributions of the massless path integral, $\A_{2,2}$ and $\A_{3,1}$, along with the respective massive improvements, which contribute to the effective action on non-trivial backgrounds and have to be projected out. It is immediate to see that the following polynomial
\begin{equation} \label{P}
P_{(3)}(z,\omega) := 4-2\left(\frac{z}{\omega}+\frac{\omega}{z}\right)-4z\omega+2\left(z^2+\omega^2 \right)= \frac{2 (z-\omega )^2 (z\omega-1)}{z\omega}
\end{equation}
does the work indeed, producing
\begin{equation}
P_{(3)}(z,\omega)\xlongrightarrow{\oint\oint} \quad Z_{\mathrm{U(1)}\times \mathrm{U(1)}} - Z_{B_{\mu\nu}} - Z_{C_{\mu}}-Z_\phi=Z_{h_{\mu\nu}}+Z_{A_{\mu}}+Z_\varphi\ .
\end{equation}
The measure \eqref{P} coincides with the one implemented to construct the path integral for the \emph{massless} graviton, although they work in different spacetime dimensions. This should not come as a surprise in retrospect: our analysis has shown that the introduction of the mass ensures that every single contribution to the partition function gets ``St\"uckelberged" while evaluating the trace on the larger Hilbert space, without the arising of unexpected terms. Consequently, the same measure eliminates both the unwanted contributions and their St\"uckelberg companions at once. \\

As discussed in \cite{Bastianelli:2019xhi}, the measure \eqref{P} can be related to a precise gauging of the R-symmetry group, specifically the gauging of the \emph{parabolic subgroup} of SO(4) (see \cite{Bastianelli:2009eh} for its application in worldline models for higher spin particles). It consists of the subgroup generated by $J_i^j$ and the trace $K^{ij}$ while excluding the insertion of the metric $G_{ij}$, which produces a measure
\begin{equation}
P_{\mathrm{par}}(z,\omega):= \frac{2 (z-\omega )^2 (z\omega-1)}{z^{3/2}\omega^{3/2}}\ .
\end{equation}
This choice leaves room for a Chern-Simons term in the Euclidean action, which can be chosen to correctly reproduce the whole measure, i.e.
\begin{equation}
S_{\mathrm{CS}}=i q\int d\tau \, a_i^i \quad \text{with} \quad q=-\frac{1}{2}\ ,
\end{equation}
which results in a modification of $P_{\mathrm{par}}(z,\omega)$ through the following multiplicative term 
\begin{equation}
P_{\mathrm{CS}}(z,\omega):=z^{1/2}\omega^{1/2} \quad \implies \quad P_{(3)}(z,\omega)=P_{\mathrm{par}}(z,\omega)P_{\mathrm{CS}}(z,\omega)\ .
\end{equation}
As previously discussed, the measure $P_{(3)}(z,\omega)$ fails in going beyond three spacetime dimensions and one has to find a way to improve it. Notably, the presence of the Chern-Simons term allows us to go to \emph{arbitrary dimensions} tuning the CS coefficient appropriately, i.e. 
\begin{equation}
q \longrightarrow q +\frac{3-D}{2}\ ,
\end{equation}
with the latter improvement producing the operatorial constraints $(\hat N_i -2)$ in arbitrary dimensions. The correct measure \eqref{P} becomes
\begin{equation} \label{PD}
P(z,\omega):=\frac{2 (z-\omega )^2 (z\omega-1)}{z\omega} \, z^{\tfrac{D-3}{2}}\omega^{\tfrac{D-3}{2}}\ ,
\end{equation}
which is of course different from the massless $D-$dimensional measure. The phase space action with the parabolic gauging reads
\begin{equation} \label{action2}
S=\int d t \left[p_\mu\dot x^{\mu}+i\bar\psi_{\mu}\cdot\dot\psi^{\mu}+i\bar \theta \cdot \dot \theta-\frac{e}{2}\,H-i\chi_i\,\bar q^i-i\bar \chi^i\, q_i-\frac12 a_{ij} K^{ij}-a_i^j(J_j^i-q\delta_j^i)\right]\ .
\end{equation}
%################################################################################################################

%################################################################################################################ 
%################################################################################################################
\section{One-loop massive gravity in the worldline formalism} \label{sec3}
In this section, we perform the quantization of the corresponding non-linear sigma model which couples a massive spin 2 particle to background gravity. It will lead to the computation of the counterterms necessary for the renormalization of the one-loop effective action of massive gravity using the worldline formalism. \\

To achieve a representation of the QFT effective action of massive gravity from \eqref{1} we need to couple the massive $\mathcal{N}=4$ spinning particle to a curved target space metric $g_{\mu\nu}(x)$. As a result, the action \eqref{action2} gets covariantized through the deformation of the worldline SUSY charges as follows
\begin{align}
\begin{split} \label{cov1}
q_i &\longrightarrow\mathcal{q}_i:=-i\,\psi_i^a\,e^\mu_a\, \pi_\mu\ +m\theta_i \\
\bar q^i &\longrightarrow\mathcal{\bar q}^i:= -i\,\bar\psi^{i\,a}\,e^\mu_a\, \pi_\mu+m\bar\theta^i
\end{split}
\end{align}
with the covariant momentum being
\begin{equation}
\pi_\mu:=p_\mu-i\omega_{\mu \, ab}\,\psi^a \cdot \bar\psi^{b}\ .
\end{equation}
Worldline fermions carry flat Lorentz indices so that $\psi^\mu_i:=e^\mu_a(x)\,\psi^a_i$, introducing a background vielbein $e_\mu^a(x)$ and the torsion-free spin connection $\omega_{\mu\, ab}$. The correct deformation of the Hamiltonian requires more consideration: the spinning particle coupled to gravity does not exhibit a first-class algebra, specifically the following anticommutator does not close:\footnote{The covariant derivatives are defined as $\hat \nabla_\mu:=\partial_\mu+\omega_{\mu\, ab}\,\psi^a\cdot\bar\psi^b$ and are related to the covariant momenta $\pi_\mu$ through $\hat \nabla_\mu=i g^{\frac{1}{4}} \pi_\mu g^{-\frac{1}{4}}$. The metric determinant factors account for a self-adjoint operator \cite{Bastianelli:2006rx}. The Laplacian is defined as
\begin{equation*}
\nabla^2:=g^{\mu\nu}\hat \nabla_\mu \hat \nabla_\nu-g^{\mu\nu}\,\Gamma^\lambda_{\mu\nu}\,\hat \nabla_\lambda\ ,
\end{equation*}
with $\Gamma^\lambda_{\mu\nu}$ being the Christoffel symbols \cite{Bastianelli:2008nm, Bonezzi:2018box}. For notational simplicity, we have used non-hermitian operators, keeping in mind that hermiticity is obtained by a similarity transformation $A \to g^{\frac{1}{4}} A g^{-\frac{1}{4}}$ on the quantum variables.}
\begin{equation}
\{\mathcal{q}_i,\mathcal{\bar q}^j \}=-\delta_i^j\left(\nabla^2-m^2\right)-R_{\mu\nu\lambda\sigma}\,\psi^\mu_i\bar\psi^{\nu\,j}\psi^\lambda\cdot\bar\psi^\sigma\ ,
\end{equation}
hence it is not immediate to identify a suitable deformed Hamiltonian. The BRST analysis of \cite{Fecit:2023kah} indicates to consider the following expression 
\begin{equation} \label{H}
H:=\nabla^2-m^2+R_{\mu\nu\lambda\sigma}\,\psi^\mu \cdot\bar\psi^\nu\psi^\lambda \cdot\bar\psi^\sigma\ .
\end{equation}
The latter Hamiltonian is necessary to achieve nilpotency of the BRST charge on the relevant physical subspace of the full BRST Hilbert space, in the \emph{massive} worldline model. Furthermore, it turns out that the background metric has to be on-shell with cosmological constant set to zero, i.e.
\begin{equation} \label{Ricci}
R_{\mu\nu}(x)=0\ .
\end{equation}
Note that within the \emph{massless} BRST quantization of \cite{Bonezzi:2018box} one has to introduce also a non-minimal coupling to the scalar curvature $\tfrac{2}{D}R$ inside the Hamiltonian \eqref{H}, since for the pure gravity case a non-zero cosmological constant is admitted. In the present case, such a coupling is inevitably zero. \\

In the following, we will evaluate perturbatively the massive path integral considering the BRST system as the starting point and keeping in mind that the results can be trusted only upon projection on Ricci-flat manifolds \eqref{Ricci}. Indeed, as previously commented, the presence of a non-trivial gravitational background obstructs the first-class character of the constraints algebra, and a more appropriate way of thinking about the model is to consider it as a genuine BRST system from the start, regardless of being derived from a gauge-invariant classical predecessor. \\

The one-loop effective action $\Gamma [g_{\mu\nu}]$ of massive gravity corresponds to the worldloop path integral of the massive $\mathcal{N}=4$ spinning particle action $S[X, G;\,g_{\mu\nu}]$, with schematic form 
\begin{equation}
\Gamma [g_{\mu\nu}]= \int_{S^1}
\frac{DG\,DX}{\mathrm{Vol(Gauge)}}\, \mathrm{e}^{-S[X,G;\,g_{\mu\nu}]}\ ,
\end{equation}
where the full action $S[X,G;\,g_{\mu\nu}]$ is the one in \eqref{action2} with the suitable covariantizations \eqref{cov1}--\eqref{H}. The gauging of the parabolic subgroup, the gauge-fixing $\tilde G=\left(T, 0, 0,\tilde a_{i}^{j}\right)$ and the path integration over the $\Phi\Pi$ ghost proceeds as outlined in the previous sections. Factorizing out the exponential of the mass, the worldloop path integral becomes 
\begin{equation} \label{euc2}
\Gamma [g_{\mu\nu}]= -\frac{1}{2}\int_{0}^{\infty}\frac{dT}{T}\,\mathrm{e}^{-m^2T} \,Z(T)
\end{equation}
where the partition function is
\begin{align} \label{PATH}
Z(T)=\frac{1}{4}
\oint \frac{dz}{2\pi i z}\oint \frac{d\omega}{2\pi i \omega}\,\frac{z}{(z+1)^2}\frac{\omega}{(\omega+1)^2} \, P(z,\omega)
\int_{_{\rm PBC}}\hskip-.4cm{ D}x\int_{_{\rm ABC}}\hskip-.4cm D\bar{\psi}D\psi\int_{_{\rm ABC}}\hskip-.4cm D\bar{\theta}D\theta \; \mathrm{e}^{-S_{\mathrm{gf}}[X,\tilde G;\,g_{\mu\nu}]}\ .
\end{align}
The gauge-fixed nonlinear sigma model action reads\footnote{To correctly define the path integral, one has to adopt a regularization scheme. In this work we exploit the calculations of \cite{Bastianelli:2023oca}, where dimensional regularization (DR) on the worldline was adopted. This choice produces a counterterm $\mathrm{V}_{\rm CT}=\frac{1}{4} R$ for the case of four worldline supersymmetries \cite{Bastianelli:2011cc}. However, such a term is vanishing upon going on-shell \eqref{Ricci}. Let us mention further that, in order to perform calculations, it is necessary to include ``metric ghosts'' to keep translational invariance of the path integral measures and to renormalize potentially divergent worldline diagrams \cite{Bastianelli:2006rx}.}
\begin{align}
\begin{split} \label{action3}
S_{\mathrm{gf}}[X,\tilde G;\,g_{\mu\nu}]&=\int d\tau \Big[ \frac{1}{4T}g_{\mu\nu}(x)\,\dot{x}^{\mu}\dot{x}^{\nu}%+m^2T
+\bar{\psi}^{a i}\left(
	\delta_i^j \mathcal{D}_\tau +i \tilde a_{i}^{j}\right){\psi}_{aj}+\bar{\theta}^{i}\left(\delta_i^j \partial_\tau +i \tilde a_{i}^{j}\right){\theta}_j \\
 &\phantom{=\int d\tau \Big[}-T R_{abcd}(x) \, \bar{\psi}^{a}\cdot \psi^{b} \bar{\psi}^{c}\cdot \psi^{d}\Big]\ ,
 \end{split}
\end{align}
where we denoted the covariant derivative with spin connection $\omega_{\mu \,ab}(x)$ acting on the fermions by
\begin{equation}
\mathcal{D}_\tau \psi^a_i := \partial_\tau \psi^a_i + \dot x^\mu \omega_{\mu}{}^a{}_b \, \psi^b_i\ .
\end{equation}
At this point, the perturbative evaluation of the path integral has to be treated with care: in particular, one has to factorize out the zero modes and expand in Riemann normal coordinates, as carefully discussed in \cite{Bastianelli:2023oca} for the massless counterpart of \eqref{PATH}. We will avoid repeating the same considerations here since the technicalities are the same. The partition function results in 
\begin{equation} \label{4}
Z(T)=\oint \frac{dz}{2\pi i}\oint \frac{d\omega}{2\pi i}\;
\mu (z,\omega) \;\int d^{D}x \, \frac{\sqrt{g(x)}}{\left(4\pi T\right)^{\frac{D}{2}}}\Big\langle \mathrm{e}^{-S_{\rm int}}\Big\rangle \ ,
\end{equation}
where the expectation value -- with normalization one, i.e. $\langle 1 \rangle = 1$, and propagators given in Appendix B of \cite{Bastianelli:2023oca} -- has to be evaluated using the Wick theorem on the free path integral, with the free action given by the quadratic part of \eqref{action3}, while higher
order terms form the interacting action $S_{\rm int}$, for which the expansion in powers of proper time $T$ has to be truncated to the desired order. In \eqref{4} we kept track inside of $\mu(z,\omega)$ of all the modular factors: these are 

\begin{enumerate}[label=(\roman*)]
 \item the parabolic measure along with the $D-$dimensional Chern-Simons term, together giving rise to $P(z,\omega)$ \eqref{PD},
 \item the $\frac{z}{(z+1)^2}\frac{\omega}{(\omega+1)^2}$ factors corresponding to the gauging of worldline supersymmetries, 
 \item the poles $\frac{1}{z}\frac{1}{\omega}$ arising from the integral measures over the moduli,
 \item the $\tfrac{1}{4}$ factor previously factored out.
 \item In addition, the path integrations over the fermionic coordinates $\int D\bar{\psi}D\psi\int D\bar{\theta}D\theta$ are responsible for the following extra factors:
\begin{align} \label{mu}
\mu (z,\omega):=\frac{1}{4}\frac{1}{(z+1)^2}\frac{1}{(\omega+1)^2}\, P(z,\omega)\; \frac{(z+1)^{D+1}}{z^{(D+1)/2}}\frac{(\omega+1)^{D+1}}{\omega^{(D+1)/2}}\ .
\end{align}
In particular, the factors $\frac{(z+1)^D}{z^{D/2}}\frac{(\omega+1)^D}{\omega^{D/2}}$ comes from the normalization of the fermionic path integral over the $\psi$s \cite{Bastianelli:2005vk}. The extra $\frac{z+1}{z^{1/2}}\frac{\omega+1}{\omega^{1/2}}$ instead is justified by the fact that the $\theta$s are free in \eqref{action3} and can be integrated over producing an additional determinant \cite{Bastianelli:2005uy}.
\end{enumerate}
The whole measure can be recast in the following form
\begin{equation}
\mu (z,\omega) =\frac{1}{2}\frac{(z+1)^{D-1}}{z^{3}}\frac{(\omega+1)^{D-1}}{\omega^{3}}(z-\omega)^{2}(z\omega-1)\ ,
\end{equation}
and corresponds to a shift $D \to D+1$ of its massless counterpart. To make explicit the Seleey-DeWitt coefficients arising from the perturbative expansion, one can introduce the double expectation value
\begin{equation}
\Big\langle \hskip -.1cm \Big\langle \mathrm{e}^{-S_{\rm int}}\Big\rangle \hskip -.1cm \Big\rangle
 = \oint \frac{dz}{2\pi i}\oint \frac{d\omega}{2\pi i}\; \mu(z,\omega) \Big\langle \mathrm{e}^{-S_{\rm int}}\Big\rangle\ .
\end{equation}
The SdW coefficients $a_{n}(D)$ parameterize the divergences as $\Big\langle \hskip -.1cm \Big\langle \mathrm{e}^{-S_{\rm int}}\Big\rangle \hskip -.1cm \Big\rangle=\sum_{n=0}^{\infty}a_{n}(D)\, T^{n}$, therefore the partition function can be rewritten as follows
\begin{equation}
Z(T)=\int d^{D}x \, \frac{\sqrt{g(x)}}{\left(4\pi T\right)^{\frac{D}{2}}} \; \sum_{n=0}^{\infty}a_{n}(D)\, T^{n}\ .
\end{equation}

%################################################################################################################ 
\subsection{One-loop divergences}
Let us start by checking the computation of the correct degrees of freedom of a massive graviton in $D$ spacetime dimensions. This is given by the double expectation value of the SdW coefficient $a_{0}(D,z,\omega)=1$. Its projected partner
\begin{equation}
a_{0}(D) = \langle \hskip -.05cm \langle 1 \rangle \hskip -.05cm \rangle = \oint \frac{dz}{2\pi i}\oint \frac{d\omega}{2\pi i}\; \mu (z,\omega) = \left. \frac{(D+1)(D-2)}{2} \right|_{D=4}= 5
\end{equation}
gives indeed the massive graviton's physical polarizations. This confirms that the measure $\mu (z,\omega)$ correctly projects only onto the degrees of freedom of a massless graviton $(h_{\mu\nu})$ plus the St\"uckelberg fields $(A_\mu,\varphi)$ introduced in order to restore gauge invariance. \\

Let us comment on the fact that if we were to insert inside \eqref{mu} the measure $P_{(3)}(z,\omega)$ without the improved $D$-dimensional Chern-Simons \eqref{P}, we would get the correct DOFs only in three spacetime dimensions while failing to go beyond that. \\

The computation of the higher-order SdW coefficients proceeds exactly as in the massless case since the $\theta$s have been integrated away. The only care one should take is to use the improved measure \eqref{mu}. With these prescriptions and with the worldline diagrams of \cite{Bastianelli:2023oca} it is possible to evaluate the counterterms up to the third order, i.e.
\begin{align} \label{series}
Z(T)=\int \frac{d^{D}x}{\left(4\pi T\right)^{\frac{D}{2}}}\sqrt{g(x)} \; \left[a_0(D)
+ a_1(D) \,T + a_2 (D)\, T^2 + a_3 (D)\, T^3+ \mathcal{O}(T^{4})\right]\ .
\end{align}
The results on Ricci-flat spaces are reported as follows. The exponentiation of all the connected diagrams and the subsequent Taylor expansion to the desired order yields
\begin{align}
\Big\langle \mathrm{e}^{-S_{\rm int}}\Big\rangle &=1+T^2\; \alpha_{2} \, R_{\mu\nu\rho\sigma}^{2} +T^3 \left( \beta_{3}\, R_{\mu\nu\rho\sigma}R^{\rho\sigma\alpha\beta}R_{\alpha\beta}{}^{\mu\nu} +\gamma_{3}\,R_{\alpha\mu\nu\beta}R^{\mu\rho\sigma\nu}R_{\rho}{}^{\alpha\beta}{}_{\sigma} \right) + \mathcal{O}(T^{4})\ .
\end{align}
where the curvature invariants of order $k$ in the Riemann tensor are multiplied by the coefficients $\alpha_{k}(z,\omega, D)$, $\beta_{k}(z,\omega, D)$ and $\gamma_{k}(z,\omega, D)$, explicitly given by \cite{Bastianelli:2023oca}
\begin{align}
\alpha_{2}&=\frac{1}{180}+\frac{1}{2} \left(\frac{\omega ^2}{(\omega +1)^4}+\frac{z^2}{(z+1)^4}+\frac{4 \omega z}{(\omega +1)^2 (z+1)^2}\right)-\frac{1}{12} \left(\frac{\omega }{(\omega+1)^2}+\frac{z}{(z+1)^2}\right)\ , \\[3mm]
\begin{split}
\beta_{3} &=\frac{17}{45360} -\frac{z}{180 (z+1)^2}-\frac{\omega }{180 (\omega +1)^2} +\frac{(z-1)^2 z^2}{6(z+1)^6} +\frac{(\omega -1)^2 \omega ^2}{6 (\omega +1)^6} \\[1.5mm]
&\phantom{=}-\frac{2 \omega z^2}{(\omega +1)^2 (z+1)^4}-\frac{2 \omega ^2 z}{(\omega +1)^4(z+1)^2}\ ,
\end{split} \\[3mm]
\gamma_{3}&=\frac{1}{1620}-\frac{z}{90(z+1)^2}-\frac{\omega }{90 (\omega +1)^2}+ \frac{z^2}{3 \left( z+1 \right)^4} + \frac{\omega^2}{3 \left( \omega+1 \right)^4} + -\frac{4 \omega ^3}{3 (\omega +1)^6}-\frac{4 z^3}{3 (z+1)^6}\ .
\end{align}
One can immediately recognize the unprojected SdW coefficients multiplying the respective powers of proper time. The final step consists of performing the modular integrals. The calculation of the various terms in the perturbative expansion delivers the following coefficients in the perturbative series \eqref{series}, including the newly found $a_3(D)$
\begin{align}
a_{0}(D)&=\frac{(D+1)(D-2)}{2}\ , \label{a0} \\[.3em]
a_{2}(D)&=\frac{D^2-31 D+508}{360} \; R_{\mu\nu\rho\sigma}^{2}\ , \label{a2} \\[.3em]
a_{3}(D)&=\frac{17 D^2-521 D-15658}{90720} \;R_{\mu\nu\rho\sigma}R^{\rho\sigma\alpha\beta}R_{\alpha\beta}{}^{\mu\nu}+\frac{D^2-37 D-1118}{3240}\; R_{\alpha\mu\nu\beta}R^{\mu\rho\sigma\nu}R_{\rho}{}^{\alpha\beta}{}_{\sigma}\ . \label{a3}
\end{align}
The expressions above are understood to be gauge-independent, as they have been calculated on Ricci-flat spaces, i.e. the background metric is on-shell. The coefficients \eqref{a0}--\eqref{a3}, including the newly computed coefficient $a_3(D)$, allow for further investigations of the issue of divergences in the quantum theory of massive gravity. The type of
divergences arising emerge naturally from the representation of the one-loop effective action with a short proper time expansion: from \eqref{series} we have
\begin{equation}
\Gamma[g_{\mu\nu}]= -\frac{1}{2}\int_{0}^{\infty}\frac{dT}{T^{1+\frac{D}{2}}} \, \mathrm{e}^{-m^2T}\int \frac{d^{D}x}{\left(4\pi \right)^{\frac{D}{2}}}\sqrt{g(x)}\left[a_0 + a_1 T + a_2 T^2 + a_3 T^3+ \mathcal{O}(T^{4})\right]\ . \label{divergences}
\end{equation} 
While the IR divergences are absent due to the presence of the mass, playing the role of a regulator, the UV divergences\footnote{To relate the $\frac{1}{\epsilon}$ pole of dimensional regularization in QFT with our result, one has to evaluate the proper time integral term by term in \eqref{divergences}, to display the gamma function dependence, as discussed in \cite{Bastianelli:2023oca}.} arise from the $T\to 0$ limit of the proper time integration. \\

In four spacetime dimensions, the different powers of T give rise to the quartic, quadratic, and logarithmic divergences parametrized by $a_0, a_1,a_2$, respectively. In QFT dimensional
regularization only the logarithmic divergences are visible. From \eqref{a2} we have
\begin{equation}
\left. a_{2} \right|_{D=4}=\frac{10}{9} \; R_{\mu\nu\rho\sigma}^{2}\ . \label{a2D=4}
\end{equation}
The latter numerical value for the one-loop four-dimensional logarithmic divergence of massive gravity coincides precisely with that calculated in \cite{Dilkes:2001av}. In four dimensions $R_{\mu\nu\rho\sigma}^{2}$ is a total derivative and can be eliminated from the effective action. Thus, one may conclude that the one-loop logarithmic divergences of massive gravity without cosmological constant vanish. More generally, the $a_2$ coefficient in $D$ dimensions has been recently evaluated in \cite{Ferrero:2023xsf} and is correctly reproduced by our result \eqref{a2}. Finally, it is worth noting that the coefficient $a_3$ gives rise to a finite term in the four-dimensional effective action. \\

In $D=6$, the coefficient $a_3$ provides an additional divergence, the logarithmic one in that dimension
\begin{equation}
\left. a_{3} \right|_{D=6} =
-\frac{649}{3240} \;R_{\mu\nu\rho\sigma}R^{\rho\sigma\alpha\beta}R_{\alpha\beta}{}^{\mu\nu}-\frac{163}{405}\; R_{\alpha\mu\nu\beta}R^{\mu\rho\sigma\nu}R_{\rho}{}^{\alpha\beta}{}_{\sigma}\ .
\end{equation}
We stress that this coefficient is gauge-independent, as the background is taken on-shell, and thus any other method of calculation should reproduce the same value. While the two curvature invariants in the previous expression are generally independent of each other, in six dimensions there exists an integral relation that connects them \cite{vanNieuwenhuizen:1976vb} and the result can be expressed as follows
\begin{align}
\left. a_{3} \right|_{D=6} = 
\frac{1}{1080} \;R_{\mu\nu\rho\sigma}R^{\rho\sigma\alpha\beta}R_{\alpha\beta}{}^{\mu\nu}\ ,
\end{align}
encoding the one-loop logarithmic divergences of massive gravity in six dimensions. 
%################################################################################################################ 

%################################################################################################################ 
%################################################################################################################

%################################################################################################################
%################################################################################################################
\section{Conclusions} \label{sec4}
In this work, we have realized the worldline path integral on the circle of the massive $\N=4$ spinning particle, providing a model capable of describing a massive graviton propagating on Ricci-flat spacetimes of arbitrary dimensions. The key point in the derivation has been to realize that the gauging of a parabolic subgroup of the SO(4) R-symmetry group, together with a suitable Chern-Simons coupling, which worked for the massless model is able to reproduce the correct result despite the mass improvement. The analysis indicates that the degrees of freedom extracted by the path integral undergo a ``St\"uckelbergization", leading to immediate identification of the massless graviton together with the associated St\"uckelberg fields, a vector and a scalar, within the spectrum of the particle model. We then applied the model to furnish a worldline representation of the effective action of massive gravity, reproducing the divergences of one-loop linearized massive gravity with vanishing cosmological constant in arbitrary dimensions. We computed the counterterms on-shell, so they furnish gauge-independent quantities which could serve as a benchmark for verifying alternative approaches to massive gravity. We checked the correct reproduction of the heat kernel coefficients $a_n(D)$ for $n=0,1,2$ comparing our results with those present in the literature. Finally, our main contribution was the determination of the Seleey-DeWitt coefficient $a_3(D)$, which to our knowledge has never been computed in the literature. \\
There are several promising directions for future exploration. One avenue involves relaxing the R-symmetry constraints to allow the propagation in the loop of the $\N=0$ supergravity, i.e. not only the graviton but also the dilaton and the Kalb-Ramond field, provided that the associated BRST system is first correctly realized, following the methodology outlined in \cite{Bonezzi:2020jjq}. Another avenue of interest lies in extending these constructions to more exotic spaces, such as non-commutative spaces, thereby realizing a covariant path integral for (spinning) particle models \cite{Bonezzi:2012vr, Franchino-Vinas:2018qyk}, or complex spaces, by employing the worldline models known as U$(\N)$ spinning particles \cite{Bastianelli:2009vj, Bastianelli:2012nh}, which may offer a fruitful first-quantized description of gravitational theories on K\"ahler manifolds once coupled to a curved background metric.
%################################################################################################################
%################################################################################################################

%################################################################################################################
%################################################################################################################
\section*{Acknowledgments}
I am grateful to F.~Bastianelli and F.~Ori for fruitful discussions. I further thank M.~M.~Cosulich for careful draft reading.
%################################################################################################################
%################################################################################################################

%################################################################################################################

%################################################################################################################
%################################################################################################################
\addcontentsline{toc}{section}{References}
\printbibliography

%################################################################################################################
%################################################################################################################

\end{document}